\documentclass[11pt,a4paper]{article}
\usepackage{graphicx}
\usepackage{amsmath}
\usepackage{amsfonts}
\usepackage{amssymb}

\begin{document}

\title{Bogoliubov transformation for distinguishable particles}
\author{M.\ Holzmann and F.\ Lalo\"{e}\\Laboratoire de Physique de l'ENS,\\LKB et LPS, 24 rue Lhomond, F-75005 Paris, France}
\maketitle
\begin{abstract}
The Bogoliubov transformation is generally derived in the context of identical
bosons with the use of ``second quantized'' $a$ and $a^{\dagger}$ operators
(or, equivalently, in field theory).\ Here, we show that the transformation,
together with its characteristic energy spectrum, can also be derived within
the Hilbert space of distinguishable particles, obeying Boltzmann statistics;
in this derivation, ordinary dyadic operators play the role usually played by
the $a$ and $a^{\dagger}$ operators; therefore, breaking the symmetry of
particle conservation is not necessary.
\end{abstract}

\bigskip

The Bogoliubov transformation \cite{Bogolubov}\cite{Bogolubov 2} is an
essential tool in the theory of Bose-Einstein condensation of identical
bosons\footnote{Historically, the first introduction of the mathematical
transformation seems to be the work of Holstein and Primakoff in 1940
\cite{Holstein}, in the context of magnetic systems.}.\ It modifies the
quadratic energy spectrum of free particles into a quasiparticle spectrum
which includes a linear variation for small momenta; this feature is generally
associated with the existence of phonons and, since it introduces a non zero
minimum value for the ratio between the energy and momentum, it allows a
natural derivation of the notion of critical velocity (a maximum velocity for
the system to remain superfluid). Usually, the mathematics of the Bogoliubov
transformation is performed within the formalism of creation and annihilation
operators (often called ``second quantization'' for historical reasons);
assuming that the system is entirely made of identical particles, one then
uses a formalism which automatically ensures a full symmetrization of the
state vector. Nevertheless, the notion of phonons has a much broader scope in
physics than just identical quantum particles; it is even often discussed in
the context of classical systems, solids or even fluids.\ One can therefore
wonder whether it is possible to re-derive the Bogoliubov spectrum in a
context where the particles are considered as distinguishable and where, as a
consequence, the effect of exchange operators remains completely
explicit.\ The purpose of the present article is to show how this is indeed
possible.\ Another motivation for a mathematical derivation of the Bogoliubov
transformation within the Hilbert space of distinguishable particles arises
for the use of Ursell operators in statistical mechanics \cite{Ursell}, a
formalism in which the symmetrization of the states is not introduced
implicitly from the beginning of the calculations, but explicitly and at a
later stage with the help of exchange cycles.

\section{Hamiltonian\label{par1}}

The hamiltonian of the problem is:%
\begin{equation}
H=\sum_{i=1}^{N}\frac{\left(  \mathbf{P}_{i}\right)  ^{2}}{2m}+\frac{1}{2}%
\sum_{i\neq j}V(i,j)\displaystyle\label{H}%
\end{equation}
where $N$ is the number of particles, $m$ their mass, $\mathbf{P}_{i}$ the
momentum of particle numbered $i$, and $V(i,j)$ the interaction energy of
particles numbered $i$ and $j$.\ The simplest assumption is to take the matrix
elements of this interaction potential as constant, provided they satisfy
momentum conservation (otherwise they of course vanish):%
\begin{equation}
<i:\mathbf{k\,\,,\,\,}j:\mathbf{k}^{^{\prime}}\mid V(i,j)\mid
i:\mathbf{k+q\,\,,\,\,}j:\mathbf{k}^{^{\prime}}-\mathbf{q}>=g \label{V}%
\end{equation}
where $g$ is the coupling constant, inversely proportional to the volume of
the system\footnote{In the thermodynamic limit, $g$ itself tends to zero, but
products such as $gN$ keep a finite value.}. The hamiltonian can then be
written:%
\begin{equation}
H=\sum_{i=1}^{N}\frac{\left(  \mathbf{P}_{i}\right)  ^{2}}{2m}+\frac{g}{2}%
\sum_{i\neq j}\sum_{\mathbf{k},\mathbf{k}^{^{\prime}},\mathbf{q}}\mid
i:\mathbf{k\,\,},\,\,j:\mathbf{k}^{^{\prime}}><i:\mathbf{k+q\,\,}%
,\,\,j:\mathbf{k}^{^{\prime}}-\mathbf{q}\mid\label{H2}%
\end{equation}

In this expression the interaction term contains, first, the forward
scattering terms $\mathbf{q}=0$ which can be written:%
\begin{equation}%
\begin{array}
[c]{l}%
\displaystyle\frac{g}{2}\sum_{i\neq j}\sum_{\mathbf{k},\mathbf{k}^{^{\prime}}%
}\mid i:\mathbf{k><}i:\mathbf{k\mid\otimes}\mid j:\mathbf{k}^{^{\prime}%
}\mathbf{><}j:\mathbf{k}^{^{\prime}}\mathbf{\mid}\\
\displaystyle\mathbf{=}\frac{g}{2}\sum_{i,j}\sum_{\mathbf{k},\mathbf{k}%
^{^{\prime}}}\mid i:\mathbf{k><}i:\mathbf{k\mid\otimes}\mid j:\mathbf{k}%
^{^{\prime}}\mathbf{><}j:\mathbf{k}^{^{\prime}}\mathbf{\mid-}\frac{g}{2}N
\end{array}
\label{V2}%
\end{equation}
or simply:%
\begin{equation}
\frac{g}{2}N(N-1)\label{V3}%
\end{equation}
As for the $\mathbf{q}\neq0$ terms, they never contain 4, or even 3, vanishing
momenta; the terms containing 2 vanishing momenta can include them, either in
the same side of the operator, or in opposite sides:%
\begin{equation}%
\begin{array}
[c]{c}%
\displaystyle\frac{g}{2}\sum_{i\neq j}\sum_{\mathbf{k}\neq0}\left[
\mid\overset{}{i}:\mathbf{0\,\,},\,\,j:\mathbf{0}><i:\mathbf{k\,\,}%
,\,\,j:-\mathbf{k}\mid+h.c.\right]  \\
\displaystyle+\frac{g}{2}\sum_{i\neq j}\sum_{\mathbf{k}\neq0}\left[
\mid\overset{}{i}:\mathbf{0\,\,},\,\,j:\mathbf{k}><i:\mathbf{k\,\,}%
,\,\,j:\mathbf{0}\mid+h.c.\right]
\end{array}
\label{v4}%
\end{equation}
where $h.c.$ is for Hermitian conjugate. It is convenient to express the
second line\footnote{For this operator in the second line, interchanging the
dummy indices $i$ and $j$ is equivalent to an hermitian conjugate operation;
for this term, we can therefore ignore the $h.c.$ and just replace $g/2$ by
$g$ .} of this expression as a function of simpler (diagonal) operators by
re-writing it in the form:%
\begin{equation}
g\sum_{i\neq j}\sum_{\mathbf{k}\neq\mathbf{0}}\left[  \mid\overset{}%
{i}:\mathbf{k\,\,},\,\,j:\mathbf{0}><i:\mathbf{k\,\,},\,\,j:\mathbf{0}%
\mid\right]  +W_{as.}\label{v5}%
\end{equation}
where $W_{as.}$ is the (antisymmetrical) interaction operator:%
\begin{equation}
W_{as.}=-g\sum_{i\neq j}\sum_{\mathbf{k}}\left[  \overset{}{1}-P_{exch.}%
(i,j)\right]  \mid i:\mathbf{k\,\,},\,\,j:\mathbf{0}><i:\mathbf{k\,\,}%
,\,\,j:\mathbf{0}\mid\label{v6}%
\end{equation}
Here, $P_{exch.}(i,j)$ is the exchange operator of particles $i$ and $j$ (the
condition $\mathbf{k}\neq\mathbf{0}$ in the summation can be released since
the corresponding term vanishes); another equivalent expression of $W_{as.}$
can be obtained by applying the exchange operator to the right\footnote{In
(\ref{v5}), (\ref{v6}) or (\ref{v6bis}), interchanging the dummy indices $i$
and $j$ is equivalent to interchanging the states labelled by $\mathbf{0}$ and
$\mathbf{k}$.}:
\begin{equation}
W_{as.}=-g\sum_{i\neq j}\sum_{\mathbf{k}}\mid i:\mathbf{0\,},\,\,j:\mathbf{k}%
><i:\mathbf{0\,\,},\,\,j:\mathbf{k}\mid\left[  \overset{}{1}-P_{exch.}%
(i,j)\right]  \label{v6bis}%
\end{equation}
(in passing, we see that $W_{as.}$ is Hermitian). Now, the first operator in
(\ref{v5}) can be simplified into:%
\begin{equation}
g\sum_{i,j}\sum_{\mathbf{k}\neq0}\left[  \mid\overset{}{i}:\mathbf{k}%
><i:\mathbf{k}\mid\otimes\mid j:\mathbf{0}><j:\mathbf{0}\mid\right]
\label{v7}%
\end{equation}
(the constraint $i\neq j$ can be released since every term $i=j$ in the
summation is zero, due to the orthogonality of single particle states), or
again:%
\begin{equation}
g\,N_{0}\sum_{\mathbf{k}\neq0}n_{\mathbf{k}}=g\,N_{0}\,N_{e}\label{v8}%
\end{equation}
where $n_{\mathbf{k}}$ is the population of the excited state labelled by
momentum $\mathbf{k}$:%
\begin{equation}
n_{\mathbf{k}}=\sum_{i}\mid i:\mathbf{k}><i:\mathbf{k}\mid\label{v9}%
\end{equation}
$N_{0}$ the population of the ground state (for the ground state, we use a
capital letter to emphasize that it has an extensive population, but $N_{0}$
is in fact the same operator as $n_{0}$):%
\begin{equation}
N_{0}=\sum_{i}\mid i:\mathbf{0}><i:\mathbf{0}\mid\label{v9bis}%
\end{equation}
$N_{e}$ as the operator associated with the total number of excited particles:%

\begin{equation}
N_{e}=\sum_{\mathbf{k\neq0}}n_{\mathbf{k}}=N-N_{0} \label{v10}%
\end{equation}
Finally, the interaction terms with 2 vanishing momenta can be written:%
\begin{equation}
\frac{g}{2}\sum_{i\neq j}\sum_{\mathbf{k}\neq0}\left[  \mid\overset{}%
{i}:\mathbf{0},j:\mathbf{0}><i:\mathbf{k},j:-\mathbf{k}\mid+h.c.\right]
+g\,N_{0}\,N_{e}+W_{as.} \label{v11}%
\end{equation}
.

Last, the only interaction term that we have not yet included corresponds to
the interaction between particles in excited states; we call this operator
$V_{ee}$:%
\begin{equation}
\displaystyle V_{ee}=\frac{g}{2}\sum_{i\neq j}\sum_{\mathbf{k},\mathbf{k}%
^{^{\prime}}\neq0}\sum_{\mathbf{q}\neq\mathbf{0},-\mathbf{k},+\mathbf{k}%
^{^{\prime}}}\left[  \mid i:\mathbf{k},j:\mathbf{k}^{^{\prime}}%
><i:\mathbf{k+q},j:\mathbf{k}^{^{\prime}}\mathbf{-q}\mid+h.c.\right]
\label{vee}%
\end{equation}
but, in what follows, we will merely neglect its effect; this is because the
system will be supposed to be at sufficiently low temperature and to be
sufficiently dilute so that most of the particles remain in the ground state;
interaction effects proportional to the square of the excited state
populations are then negligible.

To summarize, we have obtained the following expression for the Hamiltonian:%
\begin{equation}%
\begin{array}
[c]{rl}%
H & \displaystyle=\sum_{\mathbf{k}\neq0}\left(  e_{k}+gN_{0}\right)
n_{\mathbf{k}}+\frac{g}{2}N\left(  N-1\right) \\
& \displaystyle+\frac{g}{2}\sum_{i\neq j}\sum_{\mathbf{k}\neq0}\left[  \mid
i:\mathbf{0},j:\mathbf{0}><i:\mathbf{k},j:-\mathbf{k}\mid+h.c.\right] \\
& \displaystyle+V_{ee}+W_{as.}%
\end{array}
\label{Ham}%
\end{equation}
where the free particle energy $e_{k}$ is defined, as usual, by:%
\begin{equation}
e_{k}=\frac{\hslash^{2}k^{2}}{2m} \label{ec}%
\end{equation}

\section{Another expression of the hamiltonian\label{par2}}

We introduce in this section a new hamiltonian $H^{^{\prime}}$ which in a
second step, we will identify with $H$ term by term.

\subsection{Introducing new creation and annihilation operators}

We now define the operator $A_{\mathbf{k}}$ by:%
\begin{equation}%
\begin{array}
[c]{rc}%
A_{\mathbf{k}} & \displaystyle=\frac{1}{\sqrt{N}}\sum_{i}\left\{  \overset
{}{\alpha}_{\mathbf{k}}\mid i:\mathbf{0}><i:\mathbf{k}\mid+\beta_{\mathbf{k}%
}\mid i:-\mathbf{k}><i:\mathbf{0}\mid\right\}
\end{array}
\label{a}%
\end{equation}
where we assume that $\alpha_{\mathbf{k}}$ and $\beta_{\mathbf{k}}$ are real
numbers, which are for the moment not fixed, but which are supposed to be even
functions of the vector $\mathbf{k}$:%
\begin{equation}
\alpha_{\mathbf{k}}=\alpha_{-\mathbf{k}}\,\,\,\,\,;\,\,\,\,\,\,\,\beta
_{\mathbf{k}}=\beta_{-\mathbf{k}} \label{pair}%
\end{equation}
$A_{\mathbf{k}}$ is defined as an operator which removes a momentum $\hslash
k$ from the system by, either transferring one particle from momentum
$\mathbf{k}$ to zero, or from zero to $-\mathbf{k}$; it is a single particle
operator which, if restricted within the totally symmetric part $\mathcal{E}%
_{S}$ of the Hilbert space, could be expressed (through the well known
expression of single particle operators) as:
\begin{equation}
A_{\mathbf{k}}^{S}=\frac{1}{\sqrt{N}}\left\{  \alpha_{\mathbf{k}}%
a_{0}^{\dagger}a_{\mathbf{k}}+\beta_{\mathbf{k}}a_{0}a_{-\mathbf{k}}^{\dagger
}\right\}  \label{k5}%
\end{equation}
with the usual notation for the creation and annihilation operators
$a_{\mathbf{k}}$ and $a_{\mathbf{k}}^{\dagger}$.

The Hermitian conjugate of $A_{\mathbf{k}}$ is equal to:%
\begin{equation}%
\begin{array}
[c]{rc}%
A_{\mathbf{k}}^{\dagger} & \displaystyle=\frac{1}{\sqrt{N}}\sum_{j}\left\{
\overset{}{\alpha}_{\mathbf{k}}\mid j:\mathbf{k}><j:\mathbf{0}\mid
+\beta_{\mathbf{k}}\mid j:\mathbf{0}><j:-\mathbf{k}\mid\right\}
\end{array}
\label{aprime}%
\end{equation}
We can\ now calculate the commutator of $A_{\mathbf{k}}$ and $A_{\mathbf{k}%
}^{\dagger}$ term by term.\ For instance, the commutator of $\mid
i:\mathbf{0}><i:\mathbf{k}\mid$ with $\mid j:\mathbf{k}><j:\mathbf{0}\mid$ is
zero except if $i=j$; assuming that this is the case, the commutator is given
by:%
\begin{equation}
\mid i:\mathbf{0}><i:\mathbf{k}\mid i:\mathbf{k}><i:\mathbf{0}\mid-\mid
i:\mathbf{k}><i:\mathbf{0}\mid i:\mathbf{0}><i:\mathbf{k}\mid\label{abis}%
\end{equation}
which is merely the difference$\mid i:\mathbf{0}><i:\mathbf{0}\mid-\mid
i:\mathbf{k}><i:\mathbf{k}\mid$. Taking the sum of all four terms in a similar
way gives (the crossed term in $\alpha_{\mathbf{k}}\times\beta_{\mathbf{k}}$
vanish):%
\begin{equation}%
\begin{array}
[c]{rl}%
\left[  A_{\mathbf{k}}\,\,,\,A_{\mathbf{k}}^{\dagger}\right]   &
\displaystyle=N^{-1}\sum_{i}\left\{  \alpha_{\mathbf{k}}^{2}\left[  \mid
i:\overset{}{\mathbf{0}}><i:\mathbf{0}\mid-\mid i:\mathbf{k}><i:\mathbf{k}%
\mid\right]  \right.  \\
& \displaystyle\overset{}{+}\left.  \beta_{\mathbf{k}}^{2}\left[  \mid
i:-\overset{}{\mathbf{k}}><i:-\mathbf{k}\mid-\mid i:\mathbf{0}><i:\mathbf{0}%
\mid\right]  \right\}
\end{array}
\label{b}%
\end{equation}
or:%
\begin{equation}
\left[  A_{\mathbf{k}}\,\,,\,\,A_{\mathbf{k}}^{\dagger}\right]  =\left(
\alpha_{\mathbf{k}}^{2}-\beta_{\mathbf{k}}^{2}\right)  \frac{N_{0}}{N}%
+\beta_{\mathbf{k}}^{2}\,\frac{n_{-\mathbf{k}}}{N}-\alpha_{\mathbf{k}}%
^{2}\,\frac{n_{\mathbf{k}}}{N}\label{c}%
\end{equation}
or again:%
\begin{equation}
\left[  A_{\mathbf{k}}\,\,,\,A_{\mathbf{k}}^{\dagger}\right]  =\left(
\alpha_{\mathbf{k}}^{2}-\beta_{\mathbf{k}}^{2}\right)  +\beta_{\mathbf{k}}%
^{2}\,\frac{n_{-\mathbf{k}}+N_{e}}{N}-\alpha_{\mathbf{k}}^{2}\,\frac
{n_{\mathbf{k}}+N_{e}}{N}\label{e}%
\end{equation}

We now calculate the product $A_{\mathbf{k}}^{\dagger}A_{\mathbf{k}}$ for any
$\mathbf{k}\neq\mathbf{0}$; the terms $i=j$ provide:%
\begin{equation}%
\begin{array}
[c]{l}%
\displaystyle N^{-1}\sum_{i}\left[  \alpha_{\mathbf{k}}^{2}\mid\overset{}%
{i}:\mathbf{k}><i:\mathbf{k}\mid+\beta_{\mathbf{k}}^{2}\mid i:\mathbf{0}%
><i:\mathbf{0}\mid\right]  \\
\displaystyle=N^{-1}\left[  \alpha_{\mathbf{k}}^{2}\,\,n_{\mathbf{k}}%
+\beta_{\mathbf{k}}^{2}\,\,N_{0}\right]
\end{array}
\label{f}%
\end{equation}
while the terms $i\neq j$ contain a summation of the expression:%
\begin{equation}%
\begin{array}
[c]{l}%
\displaystyle N^{-1}\left\{  \alpha_{\mathbf{k}}^{2}\mid\overset{}%
{i}:\mathbf{0},\,j:\mathbf{k}><i:\mathbf{k,\,}j:\mathbf{0}\mid+\beta
_{\mathbf{k}}^{2}\mid i:-\mathbf{k},\,j:\mathbf{0}><i:\mathbf{0},\,j:-\mathbf{k}%
\mid\right.  \\
\displaystyle\left.  +\alpha_{\mathbf{k}}\times\beta_{\mathbf{k}}\left[  \mid
i:\mathbf{0},\,j:\mathbf{0}><i:\mathbf{k},\,j:-i\mathbf{k}\mid+h.\overset{\overset{}{}%
}{c}.\right]  \right\}
\end{array}
\label{g}%
\end{equation}
where, as above, $h.c.$ is for Hermitian conjugate\footnote{For this hermitian
conjugate term, we have interchanged the dummy indices $i$ and $j$.}.\ We can
now use the same method as in \S \ \ref{par1} to distinguish, within the first
line of (\ref{g}), a ``diagonal'' part and an antisymmetrical part
$B_{as.}(\mathbf{k})$:%
\begin{equation}
\frac{1}{N}\left[  \alpha_{\mathbf{k}}^{2}\,\,N_{0}\,n_{\mathbf{k}}%
+\beta_{\mathbf{k}}^{2}\,\,N_{0}\,n_{-\mathbf{k}}\right]  +B_{as.}%
(\mathbf{k})\label{h}%
\end{equation}
with the definition:%
\begin{equation}%
\begin{array}
[c]{l}%
\displaystyle B_{as.}(\mathbf{k})=-\frac{1}{N}\sum_{i\neq j}\left[
1-P_{exch.}(i,j)\right]  \left\{  \left[  \alpha_{\mathbf{k}}^{2}\mid
\overset{}{i}:\mathbf{k},\,j:\mathbf{0}><i:\mathbf{k},\,j:\mathbf{0}\mid\right.
\right.  \\
\multicolumn{1}{r}{\displaystyle\left.  +\left.  \beta_{\mathbf{k}}^{2}%
\mid\overset{}{i}:\mathbf{0},\,j:-\mathbf{k}><i:\mathbf{0},\,j:-\mathbf{k}\mid\right]
\right\}  }%
\end{array}
\label{i}%
\end{equation}
or, equivalently, by applying the exchange operator on the other side:%
\begin{equation}%
\begin{array}
[c]{l}%
\displaystyle B_{as.}(\mathbf{k})=-\frac{1}{N}\sum_{i\neq j}\left\{  \left[
\alpha_{\mathbf{k}}^{2}\mid\overset{}{i}:\mathbf{0},\,j:\mathbf{k}%
><i:\mathbf{0},\,j:\mathbf{k}\mid\right.  \right.  \\
\multicolumn{1}{r}{\displaystyle\left.  +\left.  \beta_{\mathbf{k}}^{2}%
\mid\overset{}{i}:-\mathbf{k},\,j:\mathbf{0}><i:-\mathbf{k},\,j:\mathbf{0}\mid\right]
\right\}  \left[  1-P_{exch.}(i,j)\right]  }%
\end{array}
\label{ibis}%
\end{equation}
Finally, we obtain:%
\begin{equation}%
\begin{array}
[c]{l}%
\displaystyle A_{\mathbf{k}}^{\dagger}A_{\mathbf{k}}=\alpha_{\mathbf{k}}%
^{2}\,\,\,n_{\mathbf{k}}\,(\frac{1+N_{0}}{N})+\beta_{\mathbf{k}}^{2}%
\,\,\frac{N_{0}}{N}\,\left(  1+n_{-\mathbf{k}}\right)  \\
\multicolumn{1}{r}{\displaystyle+N^{-1}\alpha_{\mathbf{k}}\,\,\beta
_{\mathbf{k}}\left\{  \sum_{i\neq j}\mid i:\mathbf{0},\,j:\mathbf{0}%
><i:\mathbf{k},\,j:-\mathbf{k}\mid+h.c.\right\}  +B_{as.}(\mathbf{k})}%
\end{array}
\label{j}%
\end{equation}

\subsection{Identification of two expressions}

Les us introduce a new hamiltonian $H^{^{\prime}}$ by:%
\begin{equation}
H^{^{\prime}}=\sum_{\mathbf{k}\neq\mathbf{0}}\widetilde{e_{k}\,}%
\,A_{\mathbf{k}}^{\dagger}A_{\mathbf{k}}+\lambda_{N}N+\lambda_{e}%
N_{e}\label{b1}%
\end{equation}
where $\alpha_{\mathbf{k}}$, $\beta_{\mathbf{k}}$, $\widetilde{e_{k}\,}$,
$\lambda_{N}$ and $\lambda_{e}$ are for the moment free parameters - in a
second step they will be chosen in order to make the new hamiltonian equal to
the initial Hamiltonian:%
\begin{equation}%
\begin{array}
[c]{rl}%
H & \displaystyle=\sum_{\mathbf{k}\neq0}\left(  e_{k}+gN_{0}\right)
n_{\mathbf{k}}+\frac{g}{2}N\left(  N-1\right)  \\
& \displaystyle+\frac{g}{2}\sum_{i\neq j}\sum_{\mathbf{k}\neq0}\left[  \mid
i:\mathbf{0},j:\mathbf{0}><i:\mathbf{k},j:-\mathbf{k}\mid+h.c.\right]
\end{array}
\label{initial}%
\end{equation}
In addition, $A_{\mathbf{k}}$ and $A_{\mathbf{k}}^{\dagger}$ will have
commutation relation that are similar to those of ordinary creation and
annihilation operators.

(i) first condition (commutation relation); if:%
\begin{equation}
\alpha_{\mathbf{k}}^{2}-\beta_{\mathbf{k}}^{2}=1 \label{b2}%
\end{equation}
(for any value of \textbf{k}), relation (\ref{e}) shows that the commutator of
$A_{\mathbf{k}}$ and $A_{\mathbf{k}}^{\dagger}$ is equal to one in the limit
of very low temperatures and very dilute systems (when almost all the
particles are in the ground state, $N_{e}\ll N$).\ Relation (\ref{b2}) is
automatically fulfilled with the following choice of the two parameters
$\alpha_{\mathbf{k}}$ and $\beta_{\mathbf{k}}$ as a function of a single
parameter $\xi_{\mathbf{k}}$:%
\begin{equation}%
\begin{array}
[c]{c}%
\alpha_{\mathbf{k}}=\cosh\,\xi_{\mathbf{k}}\\
\beta_{\mathbf{k}}=\sinh\,\xi_{\mathbf{k}}%
\end{array}
\label{c2}%
\end{equation}

(ii) second condition (identification of the main interaction terms); from
(\ref{Ham}), (\ref{j}) and (\ref{b1}) we get that the terms in $\mid
i:\mathbf{0},\,j:\mathbf{0}><i:\mathbf{k},\,j:-\mathbf{k}\mid$ can be made
identical if we set:%
\begin{equation}
\widetilde{e_{k}\,}\alpha_{\mathbf{k}}\beta_{\mathbf{k}}=\frac{gN}{2}
\label{c3}%
\end{equation}
which, through (\ref{c2}), is equivalent to:%
\begin{equation}
\widetilde{e_{k}\,}\sinh2\xi_{\mathbf{k}}=gN \label{c4}%
\end{equation}

(iii) third condition (kinetic terms in $n_{\mathbf{k}}$); the identification
of the terms which, in (\ref{j}), are linear in the excited population
operators $n_{\mathbf{k}}$ (or $n_{-\mathbf{k}}$) provides the condition:%
\begin{equation}
\widetilde{e_{k}}\left\{  \alpha_{\mathbf{k}}^{2}\frac{1+N_{0}}{N}+\left(
\beta_{-\mathbf{k}}\right)  ^{2}\frac{N_{0}}{N}\right\}  =e_{k}+gN_{0}%
\label{c5}%
\end{equation}
or, through the relation $N=N_{0}+N_{e}$:%
\begin{equation}
\widetilde{e_{k}\,}\left\{  \alpha_{\mathbf{k}}^{2}\frac{1+N-N_{e}}{N}+\left(
\beta_{-\mathbf{k}}\right)  ^{2}\frac{N-N_{e}}{N}\right\}  =e_{k}%
+gN-gN_{e}\label{c6}%
\end{equation}
In this equation, $N$ is a number while $N_{e}$ is an operator; term by term
identification then provides the two conditions:%
\begin{equation}%
\begin{array}
[c]{l}%
\widetilde{e_{k}\,}\left[  \alpha_{\mathbf{k}}^{2}(1+N^{-1})+\left(
\beta_{-\mathbf{k}}\right)  ^{2}\right]  =e_{k}+gN\\
\widetilde{e_{k}\,}\left[  \alpha_{\mathbf{k}}^{2}-\left(  \beta_{-\mathbf{k}%
}\right)  ^{2}\right]  \frac{N_{e}}{N}=gN_{e}%
\end{array}
\label{c7}%
\end{equation}
Assuming that $N\gg1$, and taking into account the parity relation
(\ref{pair}) as well as definition (\ref{c2}), we can write the former in the
form :%
\begin{equation}
\widetilde{e_{k}\,}(\alpha_{\mathbf{k}}^{2}+\beta_{\mathbf{k}}^{2}%
)=\widetilde{e_{k}\,}\cosh2\xi_{\mathbf{k}}=e_{k}+gN\label{c8}%
\end{equation}
an equality which, together with (\ref{c4}), will provide the Bogoliubov
quasiparticle spectrum. In the second line of (\ref{c7}), we have
intentionally left in both sides the operator $N_{e}$; in this way we
emphasize that, in the expression of $H^{^{\prime}}$, this term appears as a
product of $N_{e}$ by the population operator $n_{\mathbf{k}}$ of the excited
state $\mathbf{k}$, in other words as a second order correction in $N_{e}/N$
which can be neglected in the limit of low temperatures and very dilute
systems.\ Therefore, the major constraint of the identification is contained
in (\ref{c8}) and, from now on, we will leave aside the second condition of
(\ref{c7}).

(iv) terms in $N$ and $N_{e}$; in (\ref{j}), we have not yet included the
effect of the term $\left(  \beta_{\mathbf{k}}\right)  ^{2}N_{0}/N$ which,
when $N_{0}$ is replaced by $N-N_{e}$, provides:%
\begin{equation}
\frac{g}{2}N(N-1)=\sum_{\mathbf{k}}\widetilde{e_{k}\,}\,\left(  \beta
_{\mathbf{k}}\right)  ^{2}\left(  1-\frac{N_{e}}{N}\right)  +\lambda
_{N}N+\lambda_{e}N_{e}\label{c9}%
\end{equation}
Term by term identification then provides\footnote{Since, according to
(\ref{d1}), $\beta_{\mathbf{k}}$ will be proportional to $1/e_{k}$ when $k$ is
large, the sums in equation (\ref{c9}) and following are linearly divergent at
infinity. We discuss in \S \ \ref{gs} how this divergence can be eliminated.}:%
\begin{equation}
\lambda_{N}=\frac{g}{2}(N-1)-\frac{1}{N}\sum_{\mathbf{k}}\widetilde{e_{k}%
\,}\,\left(  \beta_{\mathbf{k}}\right)  ^{2}\label{c10}%
\end{equation}
and:%
\begin{equation}
\lambda_{e}=\frac{1}{N}\sum_{\mathbf{k}}\widetilde{e_{k}\,}\,\left(
\beta_{\mathbf{k}}\right)  ^{2}\label{c11}%
\end{equation}
We will neglect $\lambda_{e}$ in what follows.

(v) antisymmetric terms $B_{as.}$ and $W_{as}$; the initial hamiltonian
contains the operator $W_{as.}$ while $H^{^{\prime}}$ contains the operator:%
\begin{equation}
W_{as.}^{^{\prime}}=\sum_{\mathbf{k}}\widetilde{e_{k}\,}\,B_{as(\mathbf{k})}
\label{c12}%
\end{equation}
where $B_{as.(\mathbf{k})}$ is defined in (\ref{i}) or (\ref{ibis}).\ We note
that both these terms contain two particle antisymmetrizers $\left[
1-P_{exch.}i,j)\right]  $ and will therefore always vanish when multiplied (on
any side) by the $N$ particle symmetrization operator $S_{N}$; their
contribution is therefore exactly zero if the particles in the system are
identical bosons.\ By the same token, the same remains true if the particles
are distinguishable but in their ground state, which is also completely
symmetrical (it is actually exactly the same as for bosons).\ The property
obviously extends to any excited state having the same permutation symmetry,
but not necessarily for energy states which correspond to other
representations of the permutation group; in general, there is no reason why
the difference $W_{as.}-W_{as.}^{^{\prime}}$should play no role for
distinguishable particles\footnote{Using the parity of $\beta_{\mathbf{k}}$
and interchanging the dummy indices $i$ and $j$ in (\ref{i}) allows to show
that $W_{as.}^{^{\prime}}$ is equal to the sum over $\mathbf{k}$ of the
product $\widetilde{e_{k}\,}\,(\alpha_{\mathbf{k}}^{2}+\beta_{\mathbf{k}}%
^{2})$ by the same sum over $i\neq j$ which appears in (\ref{v6}); the strict
equality of $W_{as.}^{^{\prime}}$ and $W_{as.}$ would require the condition
$\sum_{\mathbf{k}}\widetilde{e_{k}\,}\,(\alpha_{\mathbf{k}}^{2}+\beta
_{\mathbf{k}}^{2})=gN$, which is reminiscent of the second condition
(\ref{c7}), but introduces convergence problems. Requesting a strict equality
of the operators $H$ and $H^{^{\prime}}$ in the whole Hilbert state of
Boltzmann particles would then lead to difficulties.
\par
{}}.

\bigskip

\subsection{Bogoliubov spectrum}

From (\ref{c4}) and (\ref{c8}) we get by taking the ratio:%
\begin{equation}
\tanh2\xi_{\mathbf{k}}=\frac{gN}{e_{k}+gN}\label{d1}%
\end{equation}
which fixes the value of $\xi_{\mathbf{k}}$; to determine $\widetilde{e_{k}%
\,}$ we express $\cosh2\xi_{\mathbf{k}}$ as a function of this result:%
\begin{equation}
\frac{1}{\cosh^{2}2\xi_{\mathbf{k}}}=1-\tanh^{2}2\xi_{\mathbf{k}}=\frac
{e_{k}\left(  e_{k}+2gN\right)  }{\left(  e_{k}+gN\right)  ^{2}}\label{d2}%
\end{equation}
which, combined with (\ref{c8}), gives:%
\begin{equation}
\widetilde{e_{k}\,}=\sqrt{e_{k}\left(  e_{k}+2gN\right)  }\label{d3}%
\end{equation}
This is the well known Bogoliubov result for the energy of the
quasi-particles; actually relations (\ref{d1}) and (\ref{d3}) are exactly the
basic relations obtained in the usual calculation in terms of annihilation and
creation operators.

With the above relations, the two Hamiltonians $H$ and $H^{^{\prime}}$ can be
identified, with three approximations:

* (A1) we assume that the difference $W_{as.}-W_{as.}^{^{\prime}}$ can be
ignored, which is exact for totally symmetric states.

* (A2) we ignore the effect of $V_{ee}$, corresponding to the interactions
between excited particles.

* (A3) we ignore the condition expressed in the second line of (\ref{c7}).

The third approximation is consistent with the second and remains valid at low
density and temperature, exactly as in the usual derivation for bosons.\ The
first is of different nature since it is necessary only for Boltzmann
particles; we have already mentioned that the operator $W_{as.}-W_{as.}%
^{^{\prime}}$ plays no role for the ground state of the system, as well as for
all states which are fully symmetric - for instance for all states that are
obtained from the ground state by action of any product of creation operators
$A_{\mathbf{k}}^{\dagger}$'s, since these operators do not change the
permutation symmetry of the states.

\section{Discussion}

We now discuss more precisely how the operators $A_{\mathbf{k}}$ and
$A_{\mathbf{k}}^{\dagger}$ can be used to construct eigenstates of the hamiltonian.

\subsection{Commutation relations}

One of the commutation relations between the new creation and annihilation
operators has already been studied above, and led us to an approximation which
is consistent with approximation (A2):

* (A4) we ignore, in (\ref{e}), the terms in $N_{e}/N$ (as well as those in
$n_{\mathbf{k}}/N$, but the latter tend to zero in the thermodynamic limit and
therefore raise no question).

But we also have to study the situation for different values of $\mathbf{k}$
as well as other commutations relations. We see in the definition (\ref{a}) of
$A_{\mathbf{k}}$ that its commutator with $A_{\mathbf{k}^{^{\prime}}}$ will,
first, contain the commutator:%
\begin{equation}
\left[  \mid\overset{}{i}:\mathbf{0}><i:\mathbf{k}\mid,\mid\overset{}%
{i}:\mathbf{0}><i:\mathbf{k}^{^{\prime}}\mid\right]  \label{e1}%
\end{equation}
which vanishes since both products of the operators do so (orthogonality of
single particle states if $\mathbf{k}$ and $\mathbf{k}^{^{\prime}}$ are both
different from zero). Similarly:%
\begin{equation}
\left[  \mid\overset{}{i}:-\mathbf{k}><i:\mathbf{0}\mid,\mid\overset{}%
{i}:-\mathbf{k}^{^{\prime}}><i:\mathbf{0}\mid\right]  =0\label{e2}%
\end{equation}
Two other commutators do not vanish, for instance:%
\begin{equation}%
\begin{array}
[c]{l}%
\left[  \mid\overset{}{i}:-\mathbf{k}><i:\mathbf{0}\mid,\mid\overset{}%
{i}:-\mathbf{k}^{^{\prime}}><i:\mathbf{0}\mid\right]  \\
=\mid\overset{}{i}:-\mathbf{k}><i:\mathbf{k}^{^{\prime}}\mid-\delta
_{\mathbf{k},-\mathbf{k}^{^{\prime}}}\mid\overset{}{i}:\mathbf{0}%
><i:\mathbf{0}\mid
\end{array}
\label{e3}%
\end{equation}
so that we obtain the result:%
\begin{equation}%
\begin{array}
[c]{l}%
\displaystyle\left[  A_{\mathbf{k}},A_{\mathbf{k}^{^{\prime}}}\right]
=\delta_{\mathbf{k},-\mathbf{k}^{^{\prime}}}\left(  \alpha_{\mathbf{k}%
}\,\,\beta_{-\mathbf{k}}-\alpha_{-\mathbf{k}}\,\,\beta_{\mathbf{k}}\right)
\frac{N_{0}}{N}\\
\multicolumn{1}{c}{\displaystyle-\frac{\alpha_{\mathbf{k}}\,\,\beta
_{\mathbf{k}^{^{\prime}}}}{N}\sum_{i}\mid\overset{}{i}:-\mathbf{k}^{^{\prime}%
}><i:\mathbf{k}\mid+\frac{\alpha_{\mathbf{k}^{^{\prime}}}\,\,\overset{}{\beta
}_{\mathbf{k}}}{N}\sum_{i}\mid\overset{}{i}:-\mathbf{k}><i:\mathbf{k}%
^{^{\prime}}\mid}%
\end{array}
\label{e4}%
\end{equation}
The parity relation (\ref{pair}) ensures that the main term in the right hand
side, proportional to $N_{0}/N$, vanishes exactly. The remaining terms, on the
second line, are as $V_{ee}$ ``excited-excited terms'' which act only on
excited particles, and leave them excited; it is therefore natural to
introduce one more assumption:

* (A5) the operators in the second line of (\ref{e4}) can be neglected in our
calculation, so that the $A_{\mathbf{k}}$'s get the usual commutation
relations of annihilation operators for orthogonal states.

We do not have to study the commutation relations of $A_{\mathbf{k}}^{\dagger
}$ and $A_{\mathbf{k}^{^{\prime}}}^{\dagger}$, since they can be obtained from
(\ref{e4}) by Hermitian conjugation, but we have to study the commutator of
$A_{\mathbf{k}}$ and $A_{\mathbf{k}^{^{\prime}}}^{\dagger}$.\ The calculation
from (\ref{a}) and (\ref{aprime}) is actually very similar to that which leads
to (\ref{e4}) and provides:%
\begin{equation}%
\begin{array}
[c]{l}%
\displaystyle\left[  A_{\mathbf{k}},A_{\mathbf{k}^{^{\prime}}}^{\dagger
}\right]  =\delta_{\mathbf{k},\mathbf{k}^{^{\prime}}}\left(  \alpha
_{\mathbf{k}}^{2}\,\,-\,\,\beta_{\mathbf{k}}^{2}\right)  \frac{N_{0}}{N}\\
\multicolumn{1}{c}{\displaystyle-\frac{\alpha_{\mathbf{k}}\,\,\alpha
_{\mathbf{k}^{^{\prime}}}}{N}\sum_{i}\mid\overset{}{i}:\mathbf{k}^{^{\prime}%
}><i:\mathbf{k}\mid+\frac{\,\,\beta_{\mathbf{k}}\beta_{\mathbf{k}^{^{\prime}}%
}}{N}\sum_{i}\mid\overset{}{i}:-\mathbf{k}><i:-\mathbf{k}^{^{\prime}}\mid}%
\end{array}
\label{e5}%
\end{equation}
When $\mathbf{k}=\mathbf{k}^{^{\prime}}$, this relation has already been
studied; when $\mathbf{k}\neq\mathbf{k}^{^{\prime}}$, we introduce the
additional assumption, similar to (A5):

* (A6) the operators in the second line of (\ref{e5}) can be neglected in our
calculation since they also correspond to ``excited-excited terms''.

\subsubsection{effect of the operators $A_{\mathbf{k}}$ and $A_{\mathbf{k}%
}^{\dagger}$}

For completeness, we recall here a last approximation, which has already been
discussed above and which is consistent with all preceding approximations
(valid if the gas is very dilute and at a very low temperature):

* (A7) we ignore, in the expression (\ref{b1}), the effect of the term in
$N_{e}$ (the term in $N$ creates no problem since this number is unchanged
under the action of all operators introduced; this term is studied below).

Assume then that $\mid\Phi_{E}>$ is an eigenstate of the Hamiltonian $H$, with
eigenvalue $E$; we introduce the new ket $\mid\Phi_{E}^{-}(\mathbf{k})>$ as:%
\begin{equation}
\mid\Phi_{E}^{-}(\mathbf{k})>=A_{\mathbf{k}}\mid\Phi_{E}> \label{k1}%
\end{equation}

Now, since all usual commutation relations are satisfied by the operators
$A_{\mathbf{k}}$, we can easily show that $\mid\Phi_{E}^{-}(\mathbf{k})>$ is
another eigenstate of the Hamiltonian $H$, with eigenvalue $E-\widetilde
{e_{k}\,}$.; $A_{\mathbf{k}}$ then plays the role of a ``ladder'' operator
which changes the energy step by step, by decreasing values; similarly,
$A_{\mathbf{k}}^{\dagger}$ will increase the energy eigenvalues.

Let $\mid\Phi_{0}>$ be ground state of the hamiltonian $H$.\ By action of
$A_{\mathbf{k}}$ onto this ket we obtain, either a ket which is zero, or a ket
in which the average of $H$ has decreased by $\widetilde{e_{k}\,}$, plus
possibly some corrections related to the terms in the commutators that have
been neglected.\ As noted above, $\mid\Phi_{0}>$ is a common ground state to
bosons and Boltzmann particles, since it is fully symmetric; so is
$A_{\mathbf{k}}\mid\Phi_{0}>$, which shows that the effect of $W_{as.}%
-W_{as.}^{^{\prime}}$ on this ket is strictly zero.\ We then just have to deal
with approximations (A2) to (A7), which amount to assuming that the ratio
$N_{e}/N$ has a small average value in the ground state.\ If the gas is
sufficiently dilute, all the corrections in $N_{e}/N$ will not be able to make
up for the decrease in energy $\widetilde{e_{k}\,}$ and change it into an
increase of energy.\ But no state has an average energy below that of the
ground state, so that one necessarily has:%
\begin{equation}
A_{\mathbf{k}}\mid\Phi_{0}>=0 \label{k3}%
\end{equation}
This equation is valid for any value of $\mathbf{k}$, except very small values
for which the density corrections in $N_{e}/N$ may be comparable to the
decrease in energy $\widetilde{e_{k}\,}$, so that the cancellation of the ket
is no longer a necessity. We have obtained in this way a set of equations that
the ground state of the system has to obey.

Now, applying any power of the operator $A_{\mathbf{k}}^{\dagger}$ to this
ground state will not change the permutation symmetry of the state and
therefore allow us to still ignore the effect of $W_{as.}-W_{as.}^{^{\prime}}%
$; we will therefore obtain good approximations to energy eigenstates, as long
as not too many particles are excited in the operation. The Bogoliubov
energies therefore play the role of quasiparticle energies, as in the usual
derivation for bosons, which is not surprising since the completely symmetric
subspace $\mathcal{E}_{S}$ of the Hilbert space of Boltzmann particles remains
invariant under all the operators considered.

\subsubsection{ground state energy\label{gs}}

Finally, it is interesting to come back to the term in $\lambda_{N}$ that we
have obtained in (\ref{c10}). We have:
\begin{equation}
\sum_{\mathbf{k}}\widetilde{e_{k}\,}\,\left(  \beta_{\mathbf{k}}\right)
^{2}=\sum_{\mathbf{k}}\widetilde{e_{k}\,}\,\sinh^{2}(\xi_{\mathbf{k}})
\label{d4}%
\end{equation}
where, if we use (\ref{d2}), we can insert:%
\begin{equation}
\sinh^{2}(\xi_{\mathbf{k}})=\frac{1}{2}(\cosh2\xi_{\mathbf{k}}\,-1)=\frac
{1}{2}\left\{  \frac{1}{\sqrt{e_{k}(e_{k}+2gN)}}-1\right\}  \label{d5}%
\end{equation}
which gives:%
\begin{equation}
\lambda_{N}=\frac{g}{2}(N-1)+\left\{  \sum_{\mathbf{k}}\sqrt{e_{k}(e_{k}%
+2gN)}-e_{k}-gN\right\}  \label{d6}%
\end{equation}
When $k$ tends to infinity, the expression under the sum is equivalent to:%
\begin{equation}
-\frac{(gN)^{2}}{e_{k}} \label{d7}%
\end{equation}
A cancellation of first order terms occurs, but not of second order terms, so
that the sum over all values of $\mathbf{k}$ remain (linearly) divergent.

This divergence is well known and takes place in all derivations of the
Bogoliubov transformation.\ A classical way to solve the problem
\cite{Bogolubov 2} is to replace the coupling constant $g$ (which is directly
the matrix element of the potential) by its second order expansion as a
function of the scattering length $a$:%
\begin{equation}
g=\frac{4\pi a\hslash^{2}}{mV}\left\{  1+\frac{4\pi a\hslash^{2}}{V}%
\sum_{\mathbf{k}}\frac{1}{\hslash^{2}k^{2}}+...\right\}  \label{d8}%
\end{equation}
where $V$ is the volume.\ Expressing the mean interaction term $gN(N-1)/2$ as
a function of $a$ instead of $g$ then introduced a counterterm in (\ref{d6})
which merely amounts to adding inside the sum $(gN)^{2}/e_{k}$, which pushes
the cancellation of orders up to second order, makes the sum convergent, and
proportional to the integral:%
\begin{equation}
\int_{0}^{\infty}x^{2}dx\left\{  \sqrt{x^{2}(x^{2}+2)}-x^{2}-1+\frac{1}%
{2x^{2}}\right\}  \label{d9}%
\end{equation}
The value of this integral turns out to be $\sqrt{128}/15$ and, finally, one
gets for the following well-known result for the energy of the ground state:%
\begin{equation}
\lambda_{N}N=\frac{g}{2}N(N-1)\left\{  1+\frac{128}{15}\sqrt{\frac{Na^{3}}{\pi
V}}\right\}  \label{d10}%
\end{equation}

Another method to eliminate the divergence is to use a
pseudopotential\footnote{Not to be confused with an ordinary delta function
potential; a real pseudopotential contains, in addition, a $r$ derivation
operator\cite{Huang1}\cite{Huang2} and leads to a scattering lenght which is
indeed proportional to its matrix elements, while it turns out the cross
section associated with an ordinary delta function potential vanishes.} which
directly contains the scattering length $a$ in its matrix elements, so that
any renormalization of the coupling constant such as (\ref{d8}) becomes
uncecessary. In our calculations, we have replaced all matrix elements of the
potential by the same constant $g$, but it would be possible to make a more
careful calculation with a correct expression of the matrix elements of a
pseudopotential; as shown by Castin \cite{Yvan}, the method also allows one to recover
(\ref{d10}). Beliaev \cite{Beliaev} has studied systematically how the
Bogoliubov energy spectrum with the scattering length $a$ as an interaction
parameter is recovered from a resummation of diagrams, as well as corrections
to the Bogoliubov theory. 

\begin{center}
CONCLUSION$^{{}}$
\end{center}

The mathematics of the Bogoliubov transformation can be performed within the
space of states $\mathcal{E}_{B}$ of distinguishable particles, and leads
exactly to the same formulas than in its fully symmetrical subspace
$\mathcal{E}_{S}$; in other words, the proof can be generalized to include
quantum states which are not necessarily completely symmetric by exchange and,
as one would naturally expect, renders explicit additional conditions of
validity.\ Even within the completely symmetric space $\mathcal{E}_{S}$, the
proof is not only a mathematical curiosity since it does necessitate a
symmetry breaking of the number of particles; it therefore reveals more
precisely what is behind the usual approximations that is made by replacing
the operators $a_{0}$ and $a_{0}^{\dagger}$ by c- numbers, a brutal
approximation whose effects are not necessarily easy to control
quantitatively.\ Here one gets a precise view of the exact list of operators
which have been neglected (approximations A2 to A7) - see for instance the
terms in $N_{e}/N$ in (\ref{e}) or those appearing in the second lines of
(\ref{e4}) and (\ref{e5}), which arise precisely from the fact that $a_{0}$
and $a_{0}^{\dagger}$ have not been treated as numbers; if one wished to push
the approximation beyond the usual derivation of the Bogoliubov
transformation, the consideration of the exact expression of these terms would
be useful.\ Needless to say, during the derivation, we have also obtained
trivial terms such as the interaction term between excited particles,
$V_{ee.}$, which is ignored exactly in the same way as in the traditional
derivation: it should be no surprise that, for bosons as well as for Boltzmann
particles, the ratio $N_{e}/N$ should remain small for the Bogoliubov spectrum
to be established.

In a sense, the most interesting correction term is the operator
$W_{as.}-W_{as.}^{^{\prime}}$, even if it has no effect on the whole class of
states that are common to bosons and Boltzmann particles (those which can be
obtained by the action of $A_{\mathbf{k}}^{\dagger}$ operators onto the ground
state).\ Indeed, we know the low energy spectrum of a system of
distinguishable particles is richer than that of a system of bosons, and here
this fact is reflected in our calculation by the presence of these
antisymmetric operators\footnote{In all calculations based on the Ursell
operator formalism for bosons, where a complete symmetrization is applied in a
second step, the two operators $W_{as.}$ and $W_{as.}^{^{\prime}}$ give
contributions which exactly vanish, so that the presence of these
antisymmetric operators will create no problem.}. It is well known that the
major difference between a Boltzmann system and a boson system is not to be
found in the condensate, which is basically described by exactly the same
many-body wave function in both statistics; it is in fact contained in the
excitations, which are much more numerous for Boltzmann particles since the
numbering of the particles which are excited becomes relevant.\ For
non-interacting particles, this just associates much more entropy to
excitation processes than for identical particles - this is actually the
reason behind the instability of the condensate for Boltzmann particles at any
non-zero temperature, as opposed to a Bose system where it remains stable
until the Bose Einstein temperature is reached.. As soon as interactions are
included, energetic effects also occur and, in our approach, they are
reflected by the presence of the operators $W_{as.}$ and $W_{as.}^{^{\prime}}$.

\bigskip

LKB and LPS are Unit\'{e}s associ\'{e}es au CNRS et \`{a} l'Universit\'{e}
Pierre et Marie Curie, UMR 8552 and 8551.

\bigskip
\end{document}